# Towards a terramechanics for bio-inspired locomotion in granular environments

Chen Li[1], Yang Ding[1], Nick Gravish[1], Ryan D. Maladen[2], Andrew Masse[1], Paul B. Umbanhowar[3], Haldun Komsuoglu[4], Daniel E. Koditschek[4], and Daniel I. Goldman[1]

[1]School of Physics, [2]Interdisciplinary Bioengineering Program, Georgia Institute of Technology, Atlanta, GA 30332, USA; email: {chen.li, dingyang, ngravish, rmaladen, amasse}@gatech.edu, daniel.goldman@physics.gatech.edu
[3]Department of Mechanical Engineering, Northwestern University, Evanston, IL 60208, USA; email: umbanhowar@northwestern.edu
[4]Department of Electrical and Systems Engineering, University of Pennsylvania, Philadelphia, PA 19104, USA; email: {haldunk,kod}@seas.upenn.edu

## ABSTRACT

Granular media (GM) present locomotor challenges for terrestrial and extraterrestrial devices because they can flow and solidify in response to localized intrusion of wheels, limbs, and bodies. While the development of airplanes and submarines is aided by understanding of hydrodynamics, fundamental theory does not yet exist to describe the complex interactions of locomotors with GM. In this paper, we use experimental, computational, and theoretical approaches to develop a terramechanics for bio-inspired locomotion in granular environments. We use a fluidized bed to prepare GM with a desired global packing fraction, and use empirical force measurements and the Discrete Element Method (DEM) to elucidate interaction mechanics during locomotion-relevant intrusions in GM such as vertical penetration and horizontal drag. We develop a resistive force theory (RFT) to account for more complex intrusions. We use these force models to understand the locomotor performance of two bio-inspired robots moving on and within GM.

## INTRODUCTION

Successful locomotion is crucial for space exploration. While wheeled and treaded vehicles demonstrate excellent locomotor performance and low cost of transport on rigid or close-to-rigid grounds like paved roads and hard-packed soil, the mobility of most man-made terrestrial devices is generally poor on soft, deformable substrates like granular media (GM), which are present in many terrestrial and extraterrestrial environments (Heiken et al. 1991; Matson 2010). GM present significant locomotor challenges because during intrusion of the wheels, limbs, or bodies of locomotors, GM remain solid below the yield stress, but can flow like a fluid when the yield stress is exceeded (Nedderman 1992). A relevant example is the Mars rover Spirit which became stuck in soft, loose Martian soil on several occasions (Matson 2010).

In contrast, many desert and beach dwelling animals encounter GM on a daily basis and appear to move on or even within them with ease. For example, the zebra-tailed lizard (*Callisaurus draconoides*) can run at 20 body length/s on sand (Li et al. 2011), and the sandfish lizard (*Scincus scincus*) can swim within sand at 2 body length/s (Maladen et al. 2009). Even hatchlings of aquatic sea-turtles (*Caretta caretta*) can crawl at 3 body length/s on a variety of beach sands (Mazouchova et al. 2010). Our recent biological studies have begun to reveal the mechanisms behind successful strategies used by organisms during locomotion on and within GM. For example, during running on GM, the zebra-tailed lizard paddles its large hind foot into fluidized GM to generate sufficient vertical lift to balance body weight (Li et al. 2011). During crawling on GM, the sea-turtle hatchling inserts and pushes its large flippers against solidified GM to generate sufficient thrust to overcome belly drag (Mazouchova et al. 2010). During subsurface swimming within GM, the sandfish lizard undulates its body locally fluidizing the GM to generate propulsive forces to overcome drag (Maladen et al. 2009).

Insights from biological studies promise to inspire next generation robotic devices to effectively traverse GM in terrestrial and extraterrestrial environments (Pfeifer et al. 2007). However, a major challenge to this effort remains the lack of constitutive equations for localized dense granular flow. Unlike for fluids where the interaction with flowing media can in principle be understood by solving Navier-Stokes Equations in the presence of moving boundary conditions (Vogel 1996), no fundamental force equations yet exist for GM which can be used to predict the performance of bio-inspired locomotors on GM. While continuum models are successful in cases of ideal, continuous flows such as chute flow or simple shear (Jop et al. 2006), these models cannot yet describe localized intrusion or drag of objects near granular surfaces, in which cases the flow and stress is heterogeneous and dissipative and thus may lead to complex spatiotemporal force and flow dynamics. While the field of terramechanics has been successful in modeling wheeled and tracked vehicle interaction with GM (Bekker 1960; Wong 2010), it is unclear if these models apply to the types of intrusions generated by animals or bio-inspired devices.

In this paper, we summarize our efforts using laboratory experiments, computer simulation, and theoretical modeling to begin to reveal terramechanical principles of intrusion in GM to predict bio-inspired locomotion in these environments (Ding et al. 2011b).

**GRANULAR RHEOLOGY CHANGES WITH VOLUME FRACTION**

The simplest case of intrusion into GM during bio-inspired locomotion can be modeled as either a vertical penetration or a horizontal drag. Previous studies showed that for low speed ($v < 0.5$ m/s) intrusion in GM, because the inertia of the grains is small, penetration and drag forces are nearly independent of intrusion speed, scale with the dimension of the intruder perpendicular to the direction of motion, and increase with intrusion depth (Wieghardt 1975; Albert et al. 1999; Hill et al. 2005). However, the effect of GM compaction on intrusion forces has been less explored. The compaction of GM is measured by the volume fraction, defined as the ratio between the solid volume of the medium and the volume it occupies. In nature, $\phi$ ranges from 0.58 to 0.64 (Dickinson and Ward 1994) (here we only consider dry, non-electrostatically-charged GM for which contact forces are purely repulsive). In the laboratory, ϕ can be readily and precisely controlled by a fluidized bed (Fig. 1A) (Jackson 2000). For a collection of particles, a sufficiently large upward flow of air through a distributor puts the GM into a fluidized state. As air flow is slowly turned off, grains settle into a loosely packed state (ϕ = 0.58). Subsequent air pulses or vibrations compact the GM to higher ϕ up to a maximally closely packed state (ϕ = 0.64).

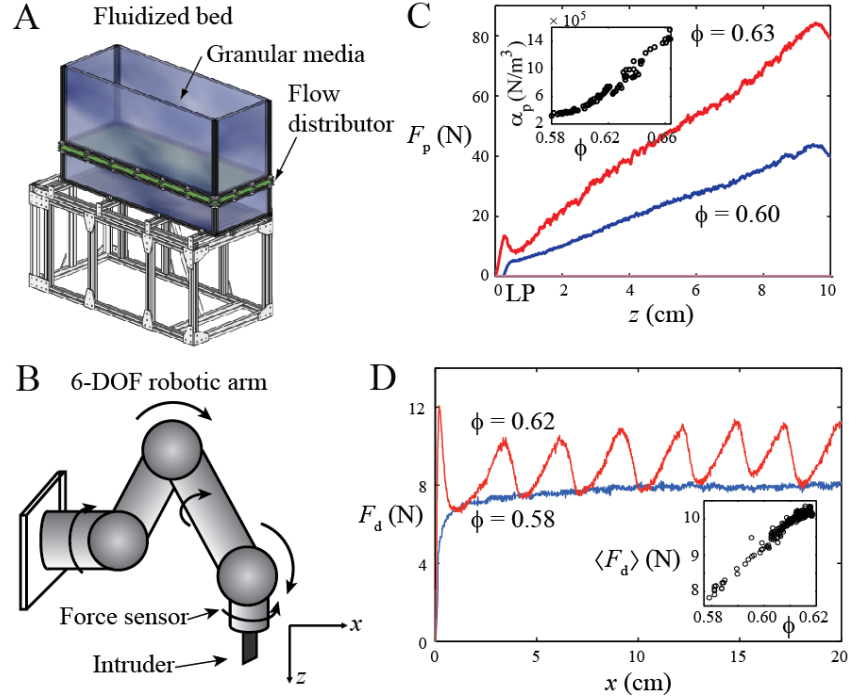

**Figure 1. Localized interaction with GM. (A) The initial ϕ is prepared by a fluidized bed. (B) A robot arm intrudes a plate while force is measured. (C) Vertical penetration force vs. penetration depth. Inset: penetration force per unit depth per unit area vs. ϕ. (D) Horizontal drag force as a function of horizontal displacement. Inset: average drag force as a function of ϕ.**

We used a 6-DOF robotic arm to move simple intruders (e.g., a plate) at low (biologically relevant) speeds (< 0.5 m/s) into GM prepared by an air fluidized bed, and investigated how penetration and drag forces depend on ϕ during slow intrusion (Fig. 1B). Vertical penetration force on a horizontally oriented plate increased approximately proportionally with plate area $A$ and penetration depth $z$, i.e., $F_p = \alpha_p A z$ (Fig. 1C) (Hill et al. 2005). $F_p$ was larger for more closely packed GM (ϕ = 0.63, red curve) than for more loosely packed GM (ϕ = 0.58, blue curve). The penetration force per unit depth per unit area $\alpha_p$ increased monotonically with ϕ (Fig. 1C, inset). The depth dependence can be understood as due to a hydrostatic-like resistance, because the force required to break frictional contacts at a given depth $z$ depends on the mass of material above, which is proportional to $z$.

Horizontal drag force on a vertically oriented plate increased approximately proportionally with plate width $w$ and approximately quadratically with maximal depth $z$ of the plate, i.e., $F_d = \alpha_d w z^2$ (Fig. 1D) (Albert et al. 1999). $F_d$ was also larger for more closely packed GM ($\phi = 0.62$, red curve) than for more loosely packed GM ($\phi = 0.58$, blue curve). The mean drag force $\langle F_d \rangle$ also increased monotonically with $\phi$ (Fig.1D, inset). The quadratic dependence on $z$ is a result of the integrating the linear hydrostatic term along the intruder. The spatiotemporal dynamics of $F_d(x)$ displayed large amplitude oscillations at high $\phi$ (Fig. 1D, red curve) (Gravish et al. 2010). The transition in $F_d$ dynamics from smooth to oscillatory occurred at the dilatancy onset ($\phi_c \approx 0.60$) (Nedderman 1992). Dilatancy is the expansion of GM under shear (Nedderman 1992). GM prepared at low $\phi$ did not dilate and exhibited a smooth drag force while GM prepared at high $\phi$ dilated during drag and periodically formed bands in which shear was concentrated which resulted in an oscillatory drag force.

## SENSITIVITY OF WALKING ON GM

To understand how $\phi$ affects locomotion on GM, we studied a insect-inspired six legged robot, SandBot (30 cm, 2.3 kg, Fig. 2A) moving on GM (~ 1 mm poppy seeds) (Li et al. 2009). SandBot has six c-shaped legs, employs an alternating tripod gait, and can bounce rapidly (2 body length/s) on solid ground. With limb kinematics adjusted to generate effective motion on GM (Li et al. 2010), SandBot walked at speeds up to 1 body length/s (30 cm/s). Speed depended sensitively the volume fraction $\phi$ of the GM (Fig. 2B). At the same stride frequency $\omega = 16$ rad/s, SandBot's speed was an order of magnitude higher on more closely packed GM ($\phi = 0.633$, red curve in Fig. 2B inset) than on more loosely packed GM ($\phi = 0.600$, blue curve in Fig. 2B inset). For sufficiently low $\omega$ and high $\omega$ (walking regime), average speed increased sublinearly with $\omega$ for any given $\phi$, and decreased with $\phi$ at a given $\omega$. However, for sufficiently high $\omega$ and/or low $\phi$ (swimming regime), average speed was consistently an order of magnitude smaller (~ 1 cm/s) (Fig. 2B).

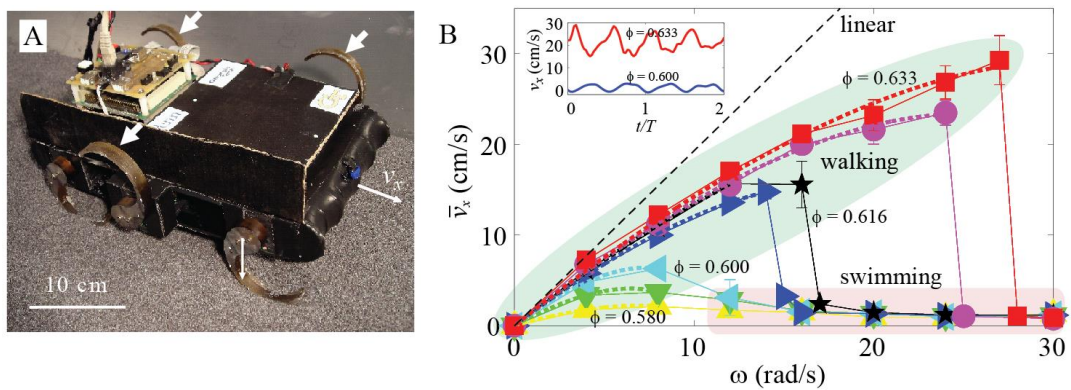

**Figure 2. Sentitivity of walking on GM. (A) SandBot standing on poppy seeds. Arrows indicate a tripod of three c-legs. (B) Average speed vs. $\omega$ for different $\phi$ ($\phi$ = 0.580, 0.590, 0.600, 0.611, 0.616, 0.622, and 0.633). Solid curves are data and dashed curves are fits from a model. Inset: speed as a function of time over two periods for two different $\phi$.**

We developed a model using the penetration force law to reveal the mechanism of effective movement of SandBot on GM (Li et al. 2009). Based on kinematic observations, we approximated the interaction of SandBot's leg with GM as vertical penetration. Because penetration force increases as $F_p = \alpha_p A z$, the legs must penetrate into the GM to a sufficient depth, $d$, to generate enough vertical force to balance the weight and inertial force of the body during locomotion. Once force was balanced, the GM under the legs solidified and the c-shaped legs started to rotate within the cavity, lifting and propeling the body forward by a step length $s$ which decreased with $d$; we refer to this as "rotary walking". As $\phi$ decreased and/or $\omega$ increased, $d$ increased and s decreased, which explained the sublinear increase of average speed because $v_x = s\omega$ in the walking regime. The model captured the observed sublinear increase of $v_x$ with $\omega$ (dash curves in Fig. 2B). As $\omega$ became high enough and/or $\phi$ low enough such that $s$ became smaller than leg length, the legs encountered previously disturbed GM over consecutive steps and dug into a hole and failed to lift and propel the body. The robot instead "swam" forward slowly (~ 1 cm/s) via frictional drag on the legs moving through fluidized GM. Thus walking on GM is most effective when kinematics are tuned to utilize the solidification features of GM.

## SWIMMING ROBOT WITHIN GM BY BODY UNDULATION

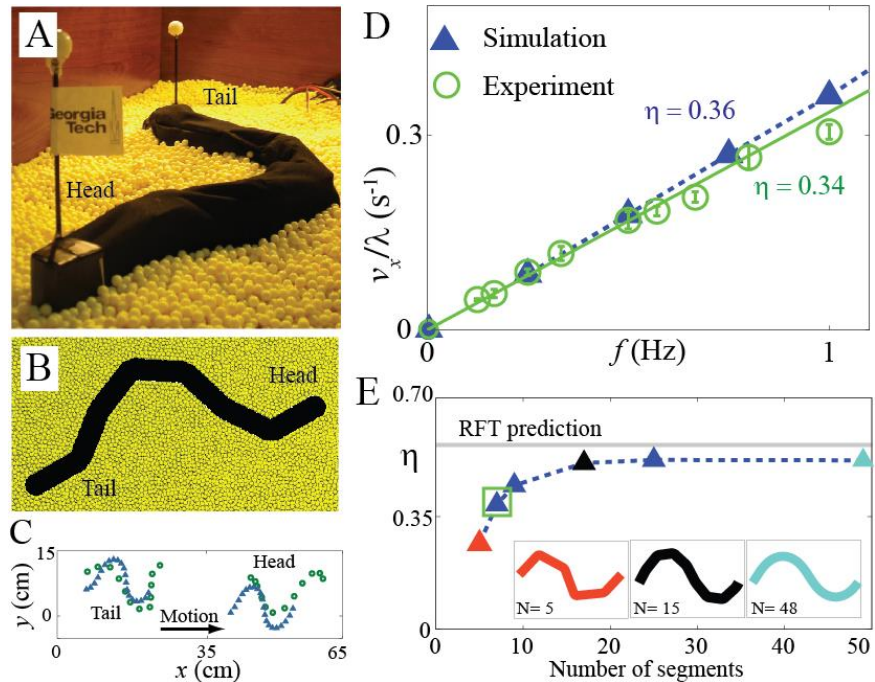

**Figure 3. (A) The sand-swimming robot sitting on 6-mm diameter plastic particles. (B) DEM simulation of the sand-swimming robot. (C) Trajectories of the head and tail of the robot from experiment (blue) and simulation (green) in the horizontal plane during sand-swimming. (D) Normalized speed as a function of frequency. (E) Wave efficiency vs. number of segments of the robot.**

Inspired by our x-ray imaging studies of the swimming of the sandfish lizard (Maladen et al. 2009), we developed in experiment and computer simulation a sand-swimming robot (Fig. 3A) capable of subsurface locomotion in GM using body undulation (Maladen et al. 2010). The robot consisted of 6 motors and a dummy wood head. The GM consisted 6mm plastic particles. The angles between the segments were controlled by the motors to generate a wave traveling posteriorly. For subsurface locomotion within GM, because both the thrust and drag forces are generated by intrusion of the locomotor and have similar dependence on $\phi$, the effect of $\phi$ on locomotor performance is less prominent (Maladen et al. 2009). Therefore we only consider an intermediate volume fraction here.

To gain insights into the interaction with GM during subsurface locomotion, we developed a computer simulation of the sand-swimming robot (Fig. 3B) (Maladen et al. 2010). The simulation coupled a numerical model of the robot created by a multi-body software to a 3D soft-sphere Discrete Element Method (DEM) (Rapaport 2004) model of the GM. In DEM simulation, interaction forces between spherical particles were modeled as Hertzian repulsive and frictional. The parameters describing the properties of the GM such as the density and friction coefficient were obtained from experiment, and the DEM model was validated by matching the resistive forces on intruders during impact and constant speed horizontal drag with experiment measurements (Maladen et al. 2011a).

Like the animal, the robot advanced forward as it undulated (Fig. 3C) subsurface; forward speed increased linearly with oscillation frequency (Fig. 3D). While simulation displayed excellent agreement with experimental measurements, the robot swam slower than the animal, as defined by its wave efficiency η, the ratio of the swimming speed to the body wave speed. The η for the animal was 0.5 while for the robot η is approximately 0.3 (Fig 3D). The simulation allowed us to study features of swimming that were inconvenient to study in robot (and biological) experiment. For example, simulation revealed that η increased with increasing number of body segments (for a fixed length device); a smoother robot achieved performance comparable to the organism (Fig. 3E). We hypothesized that the smoother body profile facilitated media flow and led to decreased drag.

**DRAG-INDUCED LIFT IN GM**

The lift forces on intruders moving horizontally within GM are relevant to locomotion on or within GM as the lift forces contribute to the vertical force to support the body during walking and can determine the vertical trajectory of the body during swimming. In both experiment and simulation, lift force $F_z$ was sensitive to the cross-sectional shape of the intruder (Fig. 4A) (Ding et al. 2011a). $F_z$ on a cylinder (left) dragged horizontally within ~ 3 mm diameter glass particles was upward, while $F_z$ on a half-cylinder with the curved surface facing upward (right) was downward but with a smaller magnitude.

Simulation revealed that the resistive force was generated primarily by the leading surface, and that the magnitude of local stress was primarily determined by the local surface orientation (Fig. 4A) (Ding et al. 2011a). The dependence of local forces on the local surface orientation suggested that decomposing the surfaces into differential area elements (plates) and summing the forces on these elements may describe the net drag and lift experienced by an intruder of complex shape. This theoretical approach is called the resistive force theory (RFT) (Gray and Hancock 1955).

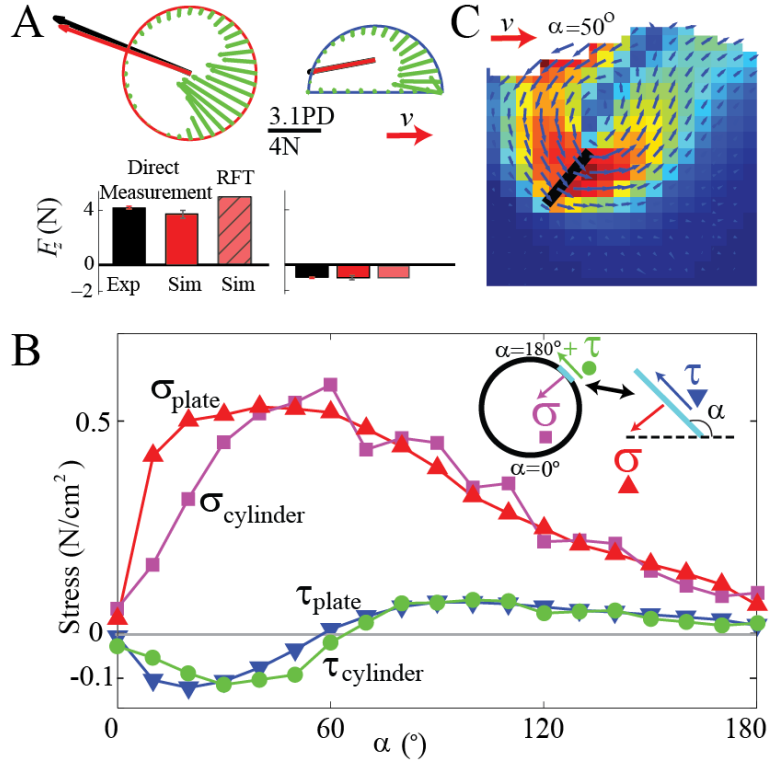

**Figure 4. Drag-induced lift force in GM. (A)** Forces on a cylinder and a half cylinder measured from experiment (black arrows and bars), simulation (red arrows and bars) and calculated from a resistive force theory (RFT) model (hatched pink bars). Green arrows indicates the forces on the surface measured from simulation. **(B)** Normal stress (σ) and shear stress (τ) on the leading surface of the cylinder as functions of tangent angle α compared to the stresses on a plate with the same α. RFT divides a complex intruder into small plates and summation of plate forces gives net lift. Inset: Correspondence between an small element on the cylinder and an inclined plate. **(C)** Flow field in the vertical plane for a representative intruder, a plate oriented at α = 50° (solid black line).

Therefore, we measured the stresses on a flat plate with tangent angle α varied between 0° and 180° (where α = 0° is along the direction of motion) (Fig. 4B) (Ding et al. 2011a). The stress on the leading side of the plate as a function of α approximately matched the stresses on a small element on cylinder oriented at the same angle α (Fig. 4B). The stresses on the plate were asymmetric about α = 90°: the normal stress increased rapidly for small α and began to decrease before reaching α = 90°. This asymmetric stress was a result of the increase of yield stress as a function of depth of the GM. The larger yield stress also resulted in an asymmetric flow field in which little flow was observed below the intruder regardless of the intruder shape (Fig. 4C). By integrating the stresses on a flat plate with tangent angle α in the range corresponding to respective shapes (e.g., $0° < \alpha < 180°$ for cylinder), lift forces on complex shape could be predicted. The RFT model indicated that the asymmetric stresses as functions of $\alpha$ were responsible for the net lift force on symmetric objects such as the cylinder. The lift force calculated from the RFT agreed well with measurements from simulation and experiment (Fig. 4A).

**VERTICAL MOTION OF SAND-SWIMMING ROBOT**

To test whether the shape-dependent lift force affects the vertical maneuvers during sand-swimming, we studied the effect of the head shape on the vertical trajectory of the sand-swimming robot (Fig. 5) (Maladen et al. 2011b). Inspired by the wedge-like head shape of the sandfish lizard, we attached wedge-shaped heads with different tilt angle $\alpha$ to the robot (Fig. 5B, inset).

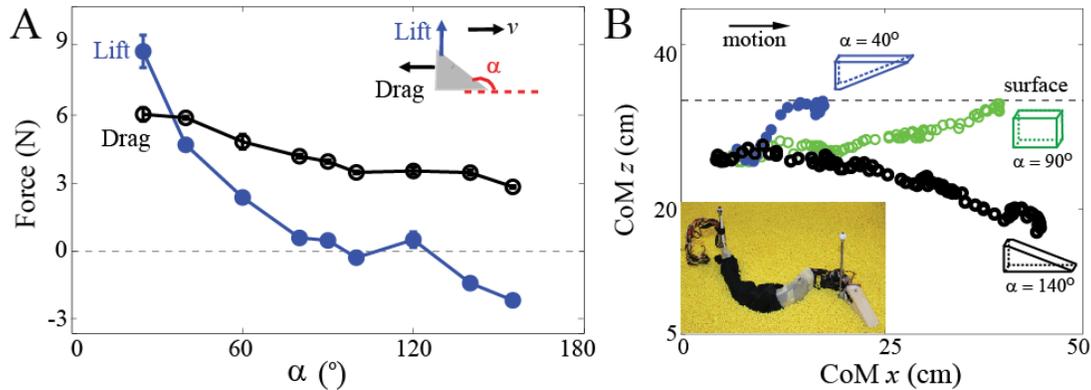

**Figure 5. Lift on different robot head shapes. (A) Lift forces on the wedge head as a function of wedge angle. Inset: schematic of experimental setup. (B) Three representative vertical-plane center of mass (CoM) trajectories with different wedge-head shapes on the robot (inset).**

When the wedges were dragged horizontally (Fig. 5A, inset), the wedges with small $\alpha$ experienced positive lift forces, whose magnitude were much larger than the those of the negative lift forces experienced by wedges with angles $\pi - \alpha$ (i.e., the same wedges flipped) (Fig. 5A, blue curve). However, the drag force was not sensitive to the shape of the intruder as is in the case of drag force in fluids (Albert et al. 2001) (Fig. 5A, black curve). When the wedges were attached to the robot (Fig. 5B, inset), the robot moved upward or downward depending on the direction of the lift force on the head Fig. 5B). The direction and the rate of the vertical motion were in agreement with the force measurements on the individual isolated heads (Fig. 5A). The robot pitched upward and rose to the surface for head shapes with $\alpha < 90°$ (blue), descended into the GM for $\alpha > 120°$ (black), and remained at approximately the same depth for $100° < \alpha < 120°$.

**CONCLUSION**

In this paper, we used laboratory experiments, computer simulation, and theoretical modeling to develop a terramechanics for bio-inspired locomotion on/within GM. We used a fluidized bed to prepare GM into a global packing fraction, and used experiments to determine empirical laws for forces as a function of volume fraction during locomotion-relevant vertical penetration and horizontal drag in GM. Both penetration and drag forces increase monotonically with volume fraction $\phi$. For $\phi$ above a critical $\phi_c$ identified as the dilation transition, a Coulomb-like yield force model of granular flow described the observed drag and lift forces. Below $\phi_c$ the force and flow dynamics were more complicated because the material underwent intermittent compaction events during flow. We observed sensitive dependence of the locomotion of a bio-inspired legged robot on $\phi$ and stride frequency, which demonstrated that effective walking on GM could be achieved by utilizing granular solidification. We developed a bio-inspired sand-swimming robot capable of moving subsurface within GM by body undulation. Using experiment and DEM simulation, we revealed the sensitive dependence of the drag induced lift on intruder shape in GM, and a RFT model predicted the lift forces on intruders of complex shapes by summing of forces on infinitesimal plates each with different orientation relative to displacement direction. The shape dependent lift forces could be utilized to control the vertical motion of the robot by adjusting the head shape of the robot.

## ACKNOWLEDGMENTS

This work was funded by the Burroughs Wellcome Fund, NSF PoLS, ARL MAST CTA, and DARPA.